\documentclass[a4paper,onecolumn,11pt,unpublished]{quantumarticle}
\pdfoutput=1
\usepackage[utf8]{inputenc}
\usepackage[english]{babel}
\usepackage[T1]{fontenc}
\usepackage{amsmath,amssymb,amsfonts}
\usepackage{graphicx}
\usepackage{booktabs}
\usepackage{multirow}
\usepackage{float}
\usepackage{algorithm}
\usepackage{algorithmic}
\usepackage{physics}
\usepackage{hyperref}
\usepackage[numbers,sort&compress]{natbib}

\begin{document}

\title{Advancing Differentiable Quantum Architecture Search via Topological Constraints, Gate Memory and Surrogate Neural Networks}

\author{David Kral}
\orcid{0009-0005-9661-2515}  
\email{dk2u25@soton.ac.uk}
\affiliation{EPSRC Centre for Doctoral Training in Quantum Technology Engineering, University of Southampton, \\SO17 1BJ, United Kingdom}
\author{Thomas Blumensath}
\affiliation{Institute of Sound and Vibration Research, University of Southampton, SO17 1BJ, United Kingdom}
\maketitle
\begin{abstract}
Quantum Architecture Search (QAS) uses classical optimisation to automatically design the parametrised circuits of Variational Quantum Algorithms. This is particularly relevant for near-term quantum hardware, where noise limits circuit depth and the optimal circuit structure for a given problem is often unknown. We propose three novel changes to Differentiable QAS (DQAS), which we call Architecture-Constrained DQAS (AC-DQAS), sequential gate Memory (M-DQAS) and their Combination (C-DQAS). In addition, we introduce surrogate-assisted variants of all four approaches (NN-DQAS, NN-AC-DQAS, NN-M-DQAS and NN-C-DQAS), which use a neural network to rank candidate circuits and simulate only those predicted to have the lowest cost. We compare these with the established ADAPT methods on three benchmarks; unitary approximation, Max-Cut and the ground-state energies of molecules. We find that AC-DQAS is the most accurate differentiable method, matching the accuracy of ADAPT-QAOA on Max-Cut while using 8 to 94 times fewer gates. For the chemistry benchmark, the four differentiable methods reach chemical accuracy in 96\% of their runs. Furthermore, the proposed methods show improved scaling in the number of circuit evaluations, growing at a rate of at most $n^{0.8}$ for systems of $n$ qubits, compared to $n^{4.8}$ to $n^{5.0}$ for the ADAPT methods. The neural network surrogate further reduces the number of circuit evaluations required during the search by factors of 12.3, 9.3 and 6.7 on average for unitary approximation, Max-Cut and molecular problems, respectively, with only a small loss in accuracy.
\end{abstract}

\section{Introduction}
\label{sec:introduction}
Quantum computation relies on manipulating quantum information encoded in complex Hilbert spaces, exploiting phenomena such as superposition and entanglement to solve certain problems more efficiently than classical computers. While theoretical breakthroughs by Shor \cite{shor_polynomial-time_1997} and Grover \cite{grover_fast_1996} have established the power of quantum algorithms, current devices are limited to the Noisy Intermediate-Scale Quantum (NISQ) regime~\cite{preskill_quantum_2018}, with $\mathcal{O}(10^2)$ to $\mathcal{O}(10^3)$ usable qubits without robust error correction. Because quantum operations are sensitive to noise, which accumulates as the circuit depth increases, long fault-tolerant algorithms remain impractical in the NISQ era \cite{preskill_quantum_2018}.

This limitation motivates the investigation of Variational Quantum Algorithms (VQAs), hybrid frameworks tailored to imperfect near-term hardware \cite{preskill_quantum_2018, cerezo_variational_2021, tilly_variational_2022}. In these methods, a parametrised quantum circuit, whose gates carry continuous parameters such as rotation angles, is trained by a classical optimiser to minimise a problem-specific cost. Prominent examples are the Variational Quantum Eigensolver (VQE) for molecular ground states \cite{peruzzo_variational_2014, tilly_variational_2022, mcardle_quantum_2020} and the Quantum Approximate Optimisation Algorithm (QAOA) for combinatorial optimisation \cite{farhi_quantum_2014}. Further applications with potential quantum advantage range from quantum simulation \cite{yuan_theory_2019} and machine learning \cite{schuld_quantum_2019} to linear systems of equations \cite{xu_variational_2021}, materials science \cite{clinton_towards_2024} and finance \cite{herman_quantum_2023}.

The performance of a VQA depends on its ansatz, in other words the structure of the parametrised circuit. A shallow ansatz may lack the expressive power to approach the solution, whereas a deep, over-parametrised one can lead to barren plateaus~\cite{mcclean_barren_2018, larocca_barren_2025}. These are regions of the optimisation landscape where gradients become exponentially small as the number of qubits increases, making them increasingly difficult to navigate. Consequently, the number of measurement shots required to estimate the gradients accurately grows exponentially as well, which can lead to the loss of any computational speed advantage that quantum computing might possess. Quantum Architecture Search (QAS) aims to avoid this by optimising the discrete structure of the circuit together with its continuous parameters, which yields compact and problem-specific circuits~\cite{martyniuk_quantum_2024, williams_automated_1999, he_training-free_2024, zhang_differentiable_2022}. The main cost then lies in the search itself, which requires evaluating a large number of candidate circuits. Unlike the number of shots needed to overcome a barren plateau, however, this number does not necessarily grow exponentially with the number of qubits. Reducing the number of circuit evaluations has therefore motivated significant research in recent years~\cite{sun_quantum_2026, liu_output_2026, zhang_neural_2021, he_gsqas_2023}.

In this paper, we improve differentiable QAS (DQAS), which learns a probability distribution over the gate at every position of the circuit by gradient descent~\cite{zhang_differentiable_2022}. Our contributions are:
\begin{itemize}
    \item Three variants of DQAS, adding an architectural constraint of alternating single- and two-qubit layers (AC-DQAS), sequential gate memory that conditions each gate on the previous ones (M-DQAS), or both (C-DQAS).
    \item Surrogate-assisted versions of all four methods (NN-DQAS, NN-AC-DQAS, NN-M-DQAS and NN-C-DQAS), in which a Surrogate Neural Network (SNN) predicts the cost of candidate circuits so that only the most promising ones are simulated.
    \item A comparison of all methods on unitary approximation, Max-Cut and the ground-state energies of molecules of 4 to 14 qubits, against the QAS methods evaluated by Lu et al.~\cite{lu_qas-bench_2023}, ADAPT-QAOA~\cite{zhu_adaptive_2022} and ADAPT-VQE~\cite{grimsley_adaptive_2019}, respectively.
\end{itemize}
We find that the constraint, alone or with memory, improves on DQAS when the operator pool consists of generic gates, as for unitary approximation and Max-Cut, but matters less when the pool is tailored to the problem, as for molecules. All differentiable methods scale far better in circuit evaluations than the ADAPT methods, with evaluations growing at most as $n^{0.8}$ with the number of qubits $n$, against $n^{4.8}$ to $n^{5.0}$. Furthermore, the SNN variants reduce these evaluations by a factor of 6.7 to 12.3 on average at a small loss in accuracy. All results come from noiseless state-vector simulations with all-to-all qubit connectivity.

The rest of this paper is structured as follows: Section~\ref{sec:Background} reviews prior work on VQAs and QAS, Section~\ref{sec:Methods} describes the developed algorithms, Section~\ref{sec:Benchmarks} introduces the benchmarks, Section~\ref{sec:Results} presents the results, and Section~\ref{sec:Discussion} discusses them, with the advantages and disadvantages of the methods and future work.

\section{Background} \label{sec:Background}

VQAs are designed to find approximate solutions to complex problems by minimising a cost (or loss) function $\mathcal{C}(\boldsymbol{\theta})$, also denoted simply by $\mathcal{C}$~\cite{cerezo_variational_2021}. A VQA begins with a fixed initial state $\ket{\psi_0}$, typically the zero state $\ket{0}^{\otimes n}$, the uniform superposition $\ket{+}^{\otimes n}$, or a problem-specific reference state. A parametrised quantum circuit applies a unitary transformation $U(\boldsymbol{\theta})$ governed by a set of continuous, trainable parameters $\boldsymbol{\theta}$. This transformation prepares a state:
\begin{equation}
    \ket{\psi(\boldsymbol{\theta})} = U(\boldsymbol{\theta})\ket{\psi_0}.
\end{equation}
To evaluate this state, a problem-specific observable $O$ is measured. The quantum computer repeatedly executes the circuit, with each execution referred to as a shot, and measures the output to estimate the expectation value:
\begin{equation}
    \langle O \rangle_{\boldsymbol{\theta}} = \bra{\psi(\boldsymbol{\theta})} O \ket{\psi(\boldsymbol{\theta})}.
\end{equation}
These expectation values are then processed by classical computers to compute $\mathcal{C}(\boldsymbol{\theta})$ and its derivative with respect to the circuit parameters \cite{wierichs_general_2022}. A classical optimiser iteratively updates $\boldsymbol{\theta}$ to minimise $\mathcal{C}(\boldsymbol{\theta})$ until convergence is achieved.

The success of a VQA depends heavily on finding an effective $U(\boldsymbol{\theta})$, that is, a suitable sequence of quantum gates and their corresponding optimal parameters. However, the most suitable gate sequence for a particular problem is generally unknown and can be searched for using QAS. Furthermore, the structure of $U(\boldsymbol{\theta})$ is constrained by the native gates available on a given quantum computing platform. We therefore assume that a predefined operator pool $\mathcal{G}$ is available, containing operations that can be implemented using individual quantum gates or combinations of them. This also allows the search to account for the target hardware, by restricting the circuit architecture to gates that are well supported by the target device, or to incorporate physical constraints, such as the conservation of electron number and spin projection in quantum chemistry (Section~\ref{sec:Bench_chem}). QAS has been extensively explored using evolutionary methods \cite{wang_quantumnas_2022, huang_robust_2022}, reinforcement learning \cite{ostaszewski_reinforcement_2021, kuo_quantum_2021, sadhu_quantum_2024}, and other machine learning techniques \cite{du_quantum_2022, su_topology-driven_2025, he_training-free_2024}.

\subsection{Greedy Methods} \label{sec:Greedy}
The first broad class of QAS methods follows a greedy search strategy. Starting from an initial ansatz, the algorithm iteratively selects and appends an operation from $\mathcal{G}$ that produces the largest immediate improvement in the cost function. This process continues until a stopping criterion is met. The circuit is therefore built incrementally by adding one operation at a time. 

A prominent example is the ADAPT-VQE algorithm for molecular ground states~\cite{grimsley_adaptive_2019}, which draws its operators from the excitation operators of the unitary coupled-cluster singles and doubles (UCCSD) ansatz (Section~\ref{sec:Bench_chem})~\cite{romero_strategies_2019}. Starting from a reference state, ADAPT-VQE evaluates the energy gradient of each operator in $\mathcal{G}$ and appends the operator with the largest gradient. After each addition, all parameters are re-optimised, and the search stops once the largest gradient falls below a predetermined threshold, indicating that additional operators would not have a significant impact on the circuit. Much of the later work focuses on changing the operator pool $\mathcal{G}$. Individual Pauli strings as operators give much shallower circuits in qubit-ADAPT-VQE~\cite{tang_qubit-adapt-vqe_2021}, while QEB-ADAPT-VQE~\cite{yordanov_qubit-excitation-based_2021} uses qubit excitations, and newer variants reduce the circuit and measurement cost further, for example with coupled exchange operators~\cite{ramoa_reducing_2025}. These improvements change the pool rather than the greedy selection, and a differentiable search could use them equally, so we focus on the original ADAPT-VQE.

The greedy rule has also been carried over to combinatorial optimisation. ADAPT-QAOA keeps the alternating layers of QAOA but chooses the mixer of each layer from $\mathcal{G}$ with the same gradient criterion, and on Max-Cut it converges much faster than the standard QAOA while needing fewer CNOT gates and parameters~\cite{zhu_adaptive_2022}.

The main advantage of greedy methods is that they do not need a fixed circuit depth, as gates are only added while they improve the cost, which tends to produce compact circuits. However, decisions based on the immediate improvement can trap the search in local minima, since added gates are never revised, and the outcome depends strongly on the initial state. More importantly for hardware, each iteration requires measuring the gradient of every operator in the pool, followed by re-optimising all circuit parameters. As a result, the number of measurements grows with both the pool size and the circuit length, while the pools themselves grow rapidly with system size~\cite{grimsley_adaptive_2019, ramoa_reducing_2025}. Each of these gradients is an expectation value that requires many measurement shots on hardware, making greedy construction expensive for larger systems. This motivates search methods whose computational cost is independent of the operator pool size or scales more favourably with it.

\subsection{Differentiable QAS} \label{sec:Differentiable}
The differentiable framework used in this paper builds upon the work of Zhang et al.~\cite{zhang_differentiable_2022}. Differentiable algorithms address the discrete nature of QAS by relaxing the architecture space into a continuous domain. A learnable weight matrix $\boldsymbol{\alpha}$ assigns continuous scores to every candidate gate from $\mathcal{G}$ at each structural position. Another tensor $\boldsymbol{\theta}$ is used for weight-sharing by tracking the expected parameter for a parametric gate at that corresponding position.

The probability of selecting a specific candidate gate $j$ at layer $l$ is obtained by applying a row-wise softmax to the architecture weights $\alpha_{l,j}$:
\begin{equation} \label{eq:softmax_methods}
    P(g_{l} = j) = \frac{\exp(\alpha_{l,j})}{\sum_{m=1}^{|\mathcal{G}|} \exp(\alpha_{l,m})},
\end{equation}
where $|\mathcal{G}|$ is the total number of operators in the pool. To optimise these weights, gradient descent is employed. Let $k \sim P(\boldsymbol{\alpha})$ represent a discrete circuit architecture sampled from this probability distribution, and let $\mathcal{L}(U(k, \boldsymbol{\theta}))$ denote the loss evaluated for that sampled circuit. The gradient of the expected cost function $\mathcal{C}$ with respect to the architecture weights is derived using the log-derivative trick~\cite{zhang_differentiable_2022}:
\begin{equation} \label{eq:policy_gradient_lit}
    \nabla_{\boldsymbol{\alpha}} \mathcal{C} = \mathbb{E}_{k \sim P} \left[ \nabla_{\boldsymbol{\alpha}} \ln P(k, \boldsymbol{\alpha}) \mathcal{L}(U(k, \boldsymbol{\theta})) \right].
\end{equation}
For a specific layer $l$ in which gate $m$ was sampled ($g_l = m$), Eq.~\ref{eq:softmax_methods} gives
\begin{align}
    \ln P(g_l = m) &= \alpha_{l,m} - \ln \sum_{r=1}^{|\mathcal{G}|} \exp(\alpha_{l,r}), \\
    \nabla_{\alpha_{l,j}} \ln P(g_l = m) &= \delta_{j,m} - P(g_l = j),
\end{align}
where $\delta_{j,m}$ is the Kronecker delta. Using these expressions, the algorithm iteratively updates the architecture weights. Gate parameters are updated one step at a time using the parameter-shift rule \cite{wierichs_general_2022}:
\begin{equation}
\label{eq:parameter_step}
\frac{\partial \mathcal{C}}{\partial \theta_p} = \frac{\mathcal{C}(\theta_p + s) - \mathcal{C}(\theta_p - s)}{2\sin(s)},
\end{equation}
where $\theta_p$ is a specific continuous parameter and $s$ is the shift, typically chosen as $\pi/2$.

Importantly, as opposed to greedy methods, the distribution over the whole circuit is optimised simultaneously, which allows the search to account for interactions between gates at different positions. Moreover, the number of circuits sampled in each iteration is set by the batch size rather than by the size of the operator pool. However, a significant issue is that the maximum circuit depth has to be fixed before the search begins. Furthermore, the number of architecture weights grows with the product of the circuit depth and the size of the operator pool, which can strain classical memory for large systems.

Differentiable search has achieved promising results across several applications. DQAS has been used for different optimisation problems~\cite{zhang_differentiable_2022}, with subsequent variants improving the search process~\cite{wu_quantumdarts_2023, sun_sa-dqas_2025}. More recently, DQAS was applied to molecules of up to 14 qubits, achieving higher accuracy with fewer CNOT gates than ADAPT-VQE when using compact circuits~\cite{chen_automated_2026}.
No previous differentiable QAS work imposes a structural template on the circuit, learns transitions between consecutive gates, or uses a surrogate model to reduce the number of circuit evaluations. These are the extensions studied in this paper.

\subsection{Topological Constraints}
Recent work suggests that the overall circuit topology has as much of an impact as the individual gate assignments \cite{su_topology-driven_2025}. Further studies have demonstrated that imposing constraints, such as alternating layers of single-qubit rotations and two-qubit entangling gates, is beneficial to circuit generation \cite{du_quantum_2022}. This naturally promotes the spread of entanglement while preventing the repetition of the same gate. Other studies have shown that circuits can be post-processed to prune unnecessary gates \cite{wang_quantumnas_2022}, or evaluated systematically to determine the impact of removing individual operations on the final output \cite{ostaszewski_structure_2021}. Collectively, these findings suggest that effective circuits exhibit certain topological patterns, which could be used to constrain the architecture search and reduce the search space.

\subsection{Surrogate Neural Networks}
Surrogate neural networks (SNNs) have been integrated into QAS to reduce the number of circuit evaluations \cite{liu_output_2026, zhang_neural_2021, he_gsqas_2023}. Early frameworks showed that neural predictors can estimate circuit performance and guide the search without direct quantum measurements \cite{zhang_neural_2021}. More recently, SNNs have been used to filter ansätze by predicting output characteristics before execution \cite{liu_output_2026}. In this hybrid approach, classical models take over most of the evaluation of the design space and reduce the number of costly and noise-prone quantum measurements.

To process these quantum architectures efficiently, predictive models must translate discrete gate arrangements into machine-readable formats. This is frequently achieved by modelling candidate circuits as Directed Acyclic Graphs (DAGs) \cite{he_gsqas_2023, he_quantum_2024, sun_quantum_2026}. The DAG representation is typically encoded via two distinct structures; an adjacency matrix detailing the connections between gates, and a feature matrix defining the specific gate assignments at each node. By feeding these two matrices into an encoder-decoder architecture, such as the Variational Graph Autoencoder (VGAE) \cite{kipf_variational_2016}, the discrete circuit topologies are mapped into a continuous latent space, which is a natural way to present this to neural networks.

\subsection{Benchmarking QAS}
Quantifying the success and efficiency of different QAS methodologies requires robust and standardised evaluation frameworks. QAS methods have been evaluated on combinatorial optimisation problems such as Max-Cut \cite{zhang_differentiable_2022, duong_quantum_2022, zhu_adaptive_2022}, ground-state energy estimation in quantum chemistry \cite{grimsley_adaptive_2019, wu_quantumdarts_2023, chen_automated_2026} and classification tasks \cite{wang_quantumnas_2022, zhang_neural_2021}, mostly with protocols defined within each study~\cite{martyniuk_quantum_2024}. In HamQASBench~\cite{niu_energy_2026}, eleven molecular Hamiltonians of up to 14 qubits are provided alongside exact reference solutions. The benchmark also shows that comparisons based on energy alone can hide meaningful differences between methods, for example in their circuit costs. We therefore report on every benchmark, alongside the accuracy, the size of the final circuit, the number of circuit evaluations of the search, and the time taken for the simulation.

A task-agnostic benchmark, QAS-Bench, introduced by Lu et al.~\cite{lu_qas-bench_2023}, focuses on unitary approximation. This dataset provides 100 target unitary matrices for systems of 2, 3, 4 and 5 qubits. Each target matrix is accompanied by training and test sets of state vectors. Each sample consists of an input vector and its corresponding output vector, generated by applying the target unitary matrix. The objective of the QAS algorithm is to recover the target unitary using only these input-output pairs. This unitary approximation benchmark provides a direct test of circuit expressivity and search efficiency. It is therefore included as one of the three benchmark tasks in this paper and is described in more detail in Section~\ref{sec:bench_unit}. However, since it does not correspond to a specific real-world application, we also include other benchmarks.

\section{Developed QAS Algorithms} \label{sec:Methods}

We developed four differentiable QAS algorithms. The first, a standard differentiable method (DQAS) following Zhang et al.~\cite{zhang_differentiable_2022}, serves as the baseline. The remaining three are newly developed in this work; an architecture-constrained variant (AC-DQAS), a variant incorporating sequential gate memory (M-DQAS) and a combined approach that integrates both architectural constraints and memory (C-DQAS). Furthermore, for each of the four methods, we developed a corresponding surrogate-assisted variant (denoted NN-DQAS, NN-AC-DQAS, NN-M-DQAS and NN-C-DQAS) that uses a surrogate neural network to predict circuit performance. As the state of the art, we compare with the greedy ADAPT methods. The hyperparameters for the differentiable methods can be found in Appendix~\ref{app:Params}.

For a given problem, we define a cost function $\mathcal{C}$ to be minimised, and a predefined pool of allowable candidate operations, denoted $\mathcal{G}$. We call it the operator pool throughout, whether its elements are single gates or operators acting on several qubits. The problem-specific operator pool, initial state and cost function of each benchmark are given in Section~\ref{sec:Benchmarks}.

\subsection{Differentiable QAS (DQAS)} \label{sec:DQAS}
DQAS follows the continuous relaxation framework introduced by Zhang et al.~\cite{zhang_differentiable_2022}, using their mathematical formulations summarised in Section~\ref{sec:Differentiable}. To this framework, we added a repetition penalty, an early stopping criterion and a final stage. The complete procedure is outlined in Algorithm~\ref{alg:diff_qas}, with the key mechanisms described below.

The search space is structured as a fixed sequence of $L$ layers (Table~\ref{tab:hyperparameters}), so the final circuit has $L$ operators. The algorithm maintains a matrix of architecture weights $\boldsymbol{\alpha}$ to compute the gate selection probabilities alongside a global weight-sharing tensor $\boldsymbol{\theta}$ that tracks the gate parameters for each respective gate position.

During the search phase, batches of candidate circuits are probabilistically sampled from the architecture distribution. For each sampled circuit, the relevant active parameters are extracted from $\boldsymbol{\theta}$, and the cost function is evaluated. This combines the standard task objective with a structural penalty that adds a problem-specific constant cost for each pair of consecutively repeated gates in the architecture, thereby discouraging the repetition of identical gates. The gradients for $\boldsymbol{\alpha}$ and $\boldsymbol{\theta}$ are computed using Eqs.~\ref{eq:policy_gradient_lit} and~\ref{eq:parameter_step}, respectively, where the mean cost of the previous batch is subtracted from each loss in Eq.~\ref{eq:policy_gradient_lit} to reduce the variance. Both $\boldsymbol{\alpha}$ and $\boldsymbol{\theta}$ are updated by one step via independent Adam optimisers~\cite{kingma_adam_2017} with learning rates $\eta_\alpha$ and $\eta_\theta$ (Table~\ref{tab:hyperparameters}). The shared parameters are initialised at random, which is described further in Section~\ref{sec:Benchmarks}.

This iterative evaluation and update process continues until an early stopping criterion is met, namely when the improvement falls below a predefined threshold $\varepsilon$. Because the global weight-sharing mechanism introduces parameter interference, the search concludes with a final stage, in one of two forms. In the sampled form, a batch of $F$ candidate architectures is drawn from the converged probability distribution, their parameters are fine-tuned for $R$ Adam steps each, and the circuit yielding the lowest final loss is returned (Table~\ref{tab:hyperparameters}). In the BFGS form, following Chen et al.~\cite{chen_automated_2026}, the most likely gate is taken in every layer, and the parameters of the resulting circuit are initialised from the search and optimised to convergence with the BFGS method~\cite{broyden_convergence_1970}. The sampled form requires the same number of circuit evaluations for every method. When comparing the scaling of individual methods, we therefore report only the evaluations required by the search itself, excluding the final stage, as this provides a clearer picture of how the search cost scales with system size. Furthermore, the final stage can easily be adapted to each task and the available computational resources.

\begin{algorithm}[!t]
\caption{Differentiable Quantum Architecture Search}
\label{alg:diff_qas}
\begin{algorithmic}[1]
\REQUIRE Operator pool $\mathcal{G}$, layers $L$, batch size $B$, max steps $S$, final samples $F$, refinement steps $R$, learning rates $\eta_\alpha, \eta_\theta$, threshold $\varepsilon$
\STATE Initialise architecture weights $\boldsymbol{\alpha} \gets \mathbf{0}$ and random shared parameters $\boldsymbol{\theta}$
\WHILE{step $t < S$}
    \STATE $P \gets \textit{ComputeProbabilities}(\boldsymbol{\alpha})$ \hfill \textit{\% Method-specific subroutine}
    \FOR{$b = 1$ \TO $B$}
        \STATE $A_b \gets \textit{SampleArchitecture}(P)$ \hfill \textit{\% Method-specific subroutine}
        \STATE Extract active parameters $\boldsymbol{\theta}_b$ from $\boldsymbol{\theta}$ corresponding to $A_b$
        \STATE Compute loss $C_b \gets \text{Evaluate}(A_b, \boldsymbol{\theta}_b) + \text{Penalty}(A_b)$
        \STATE Accumulate gradients for $\boldsymbol{\alpha}$ and $\boldsymbol{\theta}$
    \ENDFOR
    \STATE Update $\boldsymbol{\alpha}$ and $\boldsymbol{\theta}$ via Adam optimisers
    \STATE \textbf{break if} improvement is less than $\varepsilon$
\ENDWHILE
\STATE Final stage: sample $F$ architectures from $P$ and fine-tune the parameters of each for $R$ Adam steps, or take the most likely architecture and optimise its parameters with BFGS
\RETURN Circuit with minimum final loss and associated parameters
\end{algorithmic}
\end{algorithm}

\subsection{Architecture-Constrained DQAS (AC-DQAS)}
AC-DQAS enforces a strict topological constraint by constructing the circuit from blocks, each consisting of a layer of single-qubit operators applied across all qubits followed by a layer of two-qubit entangling operations, similar to the framework proposed by Du et al.~\cite{du_quantum_2022}. Integrating this constraint into differentiable QAS is novel.

To implement this, the algorithm maintains three distinct sets of continuous architecture weights instead of a single matrix, namely $\boldsymbol{\alpha}_1$ governing the single-qubit gate selection per qubit, $\boldsymbol{\alpha}_2$ governing the specific two-qubit gate selection, and $\boldsymbol{\alpha}_K$ dictating the number of entangling gates applied within each block. The maximum number of entangling gates per block is capped by the parameter $K_{\max}$ (Table~\ref{tab:hyperparameters}). These weight matrices are collectively updated using the log-derivative policy gradient shown in Eq.~\ref{eq:policy_gradient_lit}.

The core training loop remains identical to the framework outlined in Algorithm~\ref{alg:diff_qas}. The modification occurs within the specific subroutines. The \textit{ComputeProbabilities} function applies independent softmax operations to $\boldsymbol{\alpha}_1$, $\boldsymbol{\alpha}_2$ and $\boldsymbol{\alpha}_K$ to generate three distinct probability distributions. Subsequently, the \textit{SampleArchitecture} function constructs the candidate circuit appropriately to enforce the alternating topology. For each structural block, an operation is first sampled for each of the $n$ single-qubit slots, each with its own softmax and parameters, followed by sampling the integer $K$ to determine the number of entangling gates, and finally sampling exactly $K$ distinct two-qubit operations.

\subsection{Memory DQAS (M-DQAS)}
M-DQAS introduces sequential structural dependence. Rather than maintaining independent probability distributions for each fixed layer, M-DQAS models circuit generation as a sequential decision process. The algorithm defines transition weight matrices $\boldsymbol{\beta}^{(q)}$ for each individual qubit $q$. In these matrices, each row corresponds to the last gate applied to a qubit, while each column corresponds to a candidate gate that can be applied next. Each operation is listed in the matrix of every qubit it acts on; for example, a CNOT appears in the matrices of both of its qubits.

During the search phase, the algorithm tracks the active operational state of every qubit. The \textit{ComputeProbabilities} subroutine dynamically extracts the relevant rows corresponding to the current state of each qubit, concatenating them before applying a softmax function to produce a unified global probability distribution. The \textit{SampleArchitecture} subroutine then draws a single gate from this distribution, appends it to the circuit, and updates the active state tracker. This allows the algorithm to learn local patterns. Aside from these dynamic probability and sampling mechanisms, everything else remains the same, as shown in Algorithm~\ref{alg:diff_qas}. Unlike in DQAS, the number of weights in M-DQAS does not increase with circuit length. However, operations acting on multiple qubits are stored in the matrices of each qubit they act on, increasing the total number of weights. As a result, M-DQAS uses fewer weights than DQAS for long circuits, but more when the operator pool contains many multi-qubit operations. In all problems studied here, the matrices require at most 0.4~MB of memory, so classical memory is not a concern.

\subsection{Combined Constrained and Memory DQAS (C-DQAS)}
C-DQAS keeps the block structure of AC-DQAS but replaces its per-block gate weights with the transition weight matrices of M-DQAS, so that the circuit keeps the alternating topology while every gate choice depends on the gates before it. Each qubit $q$ has two matrices, $\boldsymbol{\beta}_1^{(q)}$ and $\boldsymbol{\beta}_2^{(q)}$, whose rows correspond to the last operation applied to $q$ and whose columns are the single-qubit operations and the entangling operations acting on $q$, respectively. The number of entangling gates per block is learned by $\boldsymbol{\alpha}_K$, as in AC-DQAS, up to $K_{\max}$.

\textit{SampleArchitecture} builds each block in two steps. First, every qubit draws its single-qubit operation from the row of $\boldsymbol{\beta}_1^{(q)}$ given by its last operation. Second, each entangling operation is scored by the sum of its entries in $\boldsymbol{\beta}_2^{(q)}$ over the qubits $q$ it acts on, and $K$ distinct entangling operations are drawn from the softmax over these scores, with $K$ sampled from $\boldsymbol{\alpha}_K$. After each step, the last operation applied to each qubit is updated, which is used to guide the selection of operations in the next step. All other aspects follow Algorithm~\ref{alg:diff_qas}.

\subsection{Surrogate Neural Networks} \label{sec:SNN}
The SNN needed a fixed-length description of each candidate circuit, for which we used one of two methods. The first translated generated quantum circuits into a continuous latent space. For a given problem dimension, we generated a dataset of 100,000 unique circuit architectures. Each circuit was modelled as a DAG consisting of an adjacency matrix and a feature matrix. We added the gate parameters as a column of the feature matrix, so that the model saw both the structure and the parameters.

\begin{figure}[t]
    \centering
    \begin{minipage}{0.48\textwidth}
        \centering
        \includegraphics[width=\linewidth]{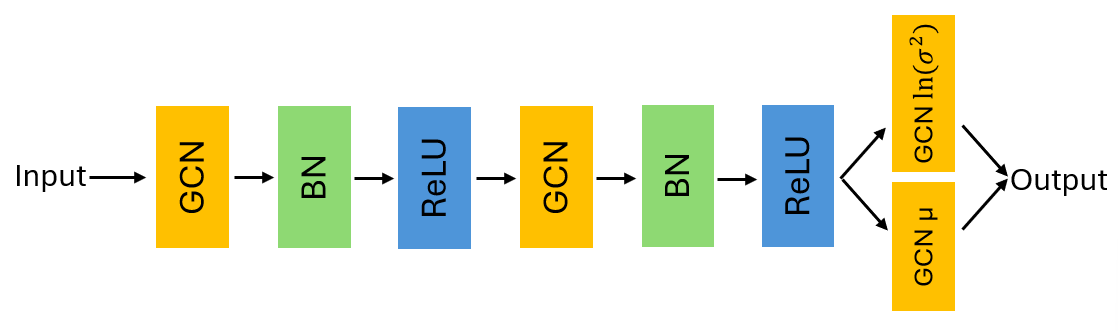}
        \vspace{-0.5em}
        \caption{Architecture of the VGAE encoder.}
        \label{fig:vgae_encoder}
    \end{minipage}\hfill
    \begin{minipage}{0.48\textwidth}
        \centering
        \includegraphics[width=\linewidth]{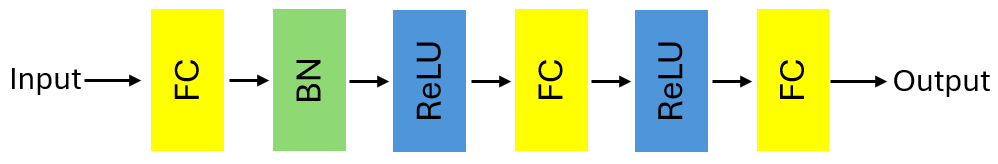}
        \vspace{-0.5em}
        \caption{Architecture of the surrogate neural network.}
        \label{fig:snn_architecture}
    \end{minipage}
    \vspace{-1em}
\end{figure}

These DAGs were mapped into a continuous domain using a VGAE, shown in Figure~\ref{fig:vgae_encoder}. The encoder had two sequential extraction blocks, each comprising a Graph Convolutional Network (GCN) layer, Batch Normalisation (BN) and a ReLU activation, followed by parallel GCN layers outputting the mean ($\mu$) and log-variance ($\ln(\sigma^2)$) of an 8-dimensional latent space. The extraction blocks used a hidden dimension of 128. The encoder and decoder were trained with the Adam optimiser using an 80/20 train-validation split and an early stopping criterion over a maximum of 500 epochs. 

The second method used the operator content of a circuit, that is, for every operator of the pool, the number of times it appears, the sum of its angles and the sum of their squares. It suits problems whose cost depends more on the parameters than on the topology, as the latent space retains the parameters poorly.

Either description was passed to the SNN, which predicted the expected cost of candidate circuits during the search. The SNN (Figure~\ref{fig:snn_architecture}) comprised a three-layer multilayer perceptron with hidden layers of 128 dimensions, using fully connected layers, BN and ReLU activations to output a single cost prediction. Ansätze generated by length-variable methods (AC-DQAS, C-DQAS) were padded with identity gates to ensure the same length.

During the primary QAS loop (Algorithm~\ref{alg:diff_qas}), the \textit{SampleArchitecture} function was modified, resulting in four distinct SNN variants: NN-DQAS, NN-AC-DQAS, NN-M-DQAS and NN-C-DQAS. At each iteration, a batch of 1024 candidate circuits was generated, encoded and evaluated by the SNN. Only the 8 candidates with the lowest predicted costs were simulated. These circuits and their costs were deposited into an experience buffer holding the last 128 batches (1024 circuits), functioning as a moving window. To combat distribution shift, the SNN was fine-tuned on this buffer for 10 epochs at each step. The primary search loop was preceded by a pre-training phase of 128 batches of 8 circuits drawn from the initial distribution to seed the predictor. The architecture update used the 8 simulated circuits in Eq.~\ref{eq:policy_gradient_lit} with the mean cost of the previous batch as the baseline, as in DQAS, with the learning rates of each SNN variant given in Table~\ref{tab:hyperparameters}.

The SNN reduces the number of circuit evaluations at the cost of extra classical computation. Its main storage cost is the VGAE training set, which takes 0.1 to 1.2~GB per problem size. This set was generated and trained on only once and reused for every instance of that size. At every step, 1024 candidates also had to be generated and encoded, and the SNN fine-tuned on its buffer. At the largest size of each benchmark, this took less time than it saved by simulating fewer circuits for every variant except NN-M-DQAS and, on the unitaries, NN-C-DQAS (Appendix~\ref{app:Time}).

\subsection{ADAPT-VQE and ADAPT-QAOA} \label{sec:ADAPT}
For comparison, we use the greedy methods ADAPT-VQE~\cite{grimsley_adaptive_2019} for the molecular benchmark and ADAPT-QAOA~\cite{zhu_adaptive_2022} for the Max-Cut benchmark. Starting from the initial state, ADAPT-VQE computes the energy gradient of every operator in $\mathcal{G}$ at zero angle, appends the operator with the largest gradient magnitude, re-optimises all parameters with BFGS (gradient tolerance $10^{-7}$), and stops when the norm of the gradient vector falls below $10^{-3}$. We validated our implementation against the authors' code and confirmed that, using the original operator pool, it reproduces the published ansätze~\cite{grimsley_adaptive_2019}.

ADAPT-QAOA starts from the initial state $\ket{+}^{\otimes n}$ and builds the circuit by adding layers of the form $\exp(-i\beta_k A_k)\exp(-i\gamma_k H_C)$, where $H_C$ is the cost Hamiltonian of the problem. At each step, the mixer operator $A_k$ is selected as the operator with the largest energy gradient from $\mathcal{G}$. The gradients are evaluated after applying a trial cost layer with $\gamma_k = 0.01$. All parameters are re-optimised with BFGS after every layer, and the search stops when the largest improvement falls below $10^{-3}$. We checked our implementation against the open-source code released with the Dynamic-ADAPT-QAOA study of Yanakiev et al.~\cite{yanakiev_dynamic_2024}, which gives the same pool gradients and energies. Every ADAPT-QAOA layer contains the full cost Hamiltonian, so its circuits can be much larger than those of the differentiable methods. We therefore report it both as published and stopped after its first layer, the smallest circuit it can return.

\section{Experimental Benchmarks} \label{sec:Benchmarks}
To assess their effectiveness, the algorithms were evaluated on three benchmark tasks spanning different areas of quantum computing: unitary approximation, Max-Cut and molecular ground-state estimation.

Each differentiable method and its variant was run once on each of the 100 instances per size of unitary approximation and Max-Cut, and three times, with different random seeds, on each molecular case. The hyperparameters for every method and each benchmark, including the circuit size, the learning rates and $K_{\max}$, are listed in Table~\ref{tab:hyperparameters}.

\subsection{Unitary Approximation}\label{sec:bench_unit}
The first benchmark evaluates the algorithms on unitary approximation, using the dataset introduced by Lu et al.~\cite{lu_qas-bench_2023}, which has 100 problems for 2, 3, 4 and 5 qubits each. The objective is to reconstruct an unknown target unitary $U^t$. The cost function is the sum of the absolute differences between the matrix elements of the unitary $U^s$ of the current candidate ansatz and those of $U^t$:
\begin{equation}
\label{eq:matrix_comp}
\mathcal{C} = \sum_{i=0}^{2^n-1}\sum_{j=0}^{2^n-1} \left| U^s_{ij} - U^t_{ij} \right|.
\end{equation}
After training, the final circuit is scored on the dataset's input and output vectors by the state fidelity metric $f$, which quantifies the similarity between the predicted output state $\tilde{\psi}$ and the target state $\psi$:
\begin{equation} \label{eq:f}
f = \left( \sum_{j=0}^{2^n-1} \sqrt{a_j^2 + b_j^2}\,\sqrt{\hat{a}_j^2 + \hat{b}_j^2} \right)^2,
\end{equation}
where $a, b$ and $\hat{a}, \hat{b}$ represent the real and imaginary coefficients of the respective basis states. Here, $\psi = U^t \ket{x}$ is the output of the target unitary for an input state $\ket{x}$ from the dataset's test set. As $f \to 1$, the predicted and target states yield identical measurement probability distributions. The search thus minimises Eq.~\ref{eq:matrix_comp}, while the final circuits are scored by $f$.

The operator pool consists of parametrised single-qubit rotations ($R_X, R_Y, R_Z$) applied to any individual qubit, and non-parametric two-qubit entangling gates (CNOT) applied between all possible pairs of qubits. The initial rotation angles are drawn uniformly from the interval $[0, 2\pi)$, the repetition penalty is $10^{-1}$, the search ends with the sampled form of the final stage ($F = 16$, $R = 300$; 4800 circuit evaluations for every method) and the SNN receives the latent space of the VGAE. 

Unitary approximation has no established greedy method, so we compare against the QAS methods evaluated by Lu et al.~\cite{lu_qas-bench_2023} on the same dataset. These are a reinforcement-learning agent that adds one gate at a time (RL)~\cite{ostaszewski_reinforcement_2021}, DQAS (DA)~\cite{zhang_differentiable_2022} and the hybrid method of Du et al.~\cite{du_quantum_2022} (HA). HA samples layered circuits with a rotation on every qubit in each layer, optimises their parameters for a few steps and fine-tunes the best candidate. For the 4- and 5-qubit cases, Lu et al. trained DA using a softmax-weighted mixture of all gates rather than sampled circuits~\cite{lu_qas-bench_2023}. However, since the circuit lengths and other relevant parameters are not reported, a direct comparison is not possible, and we can only use their reported results as a reference.

\subsection{Max-Cut Combinatorial Optimisation}
The second benchmark investigates the Max-Cut combinatorial optimisation problem, which requires partitioning the nodes of a graph into two disjoint sets such that the number of edges connecting the sets is maximised. It is a widely studied problem in combinatorial optimisation and it is often the focus of QAS papers~\cite{zhu_adaptive_2022, yanakiev_dynamic_2024, zhang_differentiable_2022}. This problem is mapped to a cost function using a QAOA-style cost Hamiltonian:
\begin{equation}
H_C = \frac{1}{2} \sum_{(i,j) \in E} \left( I - Z_i Z_j \right),
\end{equation}
where $E$ represents the set of edges, and $Z_i$ and $Z_j$ denote the Pauli Z operators acting on qubits $i$ and $j$, respectively~\cite{zhang_differentiable_2022, duong_quantum_2022}. Its expectation value is the expected cut of the state, so the differentiable methods minimise $\mathcal{C} = \sum_{(i,j) \in E} \langle Z_i Z_j \rangle = |E| - 2\langle H_C \rangle$, which maximises the expected cut. This is an Ising-type cost function, a common form in VQA problems. Many hard combinatorial problems can be encoded in this form~\cite{lucas_ising_2014}. To ensure a robust evaluation, we generated 100 distinct unweighted Max-Cut instances for each of 8, 12 and 16 nodes (qubits), in which every edge is present with probability 0.5. Each graph was resampled until connected and checked to be non-isomorphic to all others, to prevent trivial or redundant instances, and its maximum cut was computed exactly. The differentiable methods use the same operator pool ($R_X, R_Y, R_Z$, CNOT) and final stage settings as for unitary approximation. Unlike for unitary approximation, the circuits start from $\ket{+}^{\otimes n}$, prepared by $n$ Hadamard gates. As the greedy baseline, we run ADAPT-QAOA (Section~\ref{sec:ADAPT}), whose cost layers apply three gates per edge of the graph.

\subsection{Molecular Ground States} \label{sec:Bench_chem}
The last benchmark is finding the ground-state energy of small molecules, which is the main application of VQE and one of the most promising uses of near-term quantum computers~\cite{peruzzo_variational_2014, mcardle_quantum_2020}. The ground-state energy, the lowest energy of the electrons for fixed positions of the nuclei, determines properties such as bond lengths and reaction energies, but its exact computation scales exponentially with the size of the molecule~\cite{mcardle_quantum_2020}. Using the Born-Oppenheimer approximation, the electronic Hamiltonian is written in second quantisation as
\begin{equation} \label{eq:H_el}
    H = \sum_{p,q} h_{pq}\, a_p^\dagger a_q + \frac{1}{2} \sum_{p,q,r,s} h_{pqrs}\, a_p^\dagger a_q^\dagger a_r a_s,
\end{equation}
where $a_p^\dagger$ and $a_p$ are the creation and annihilation operators of an electron in spin orbital $p$, a one-electron state with a given spin, and the one- and two-electron integrals $h_{pq}$ and $h_{pqrs}$ are computed classically~\cite{mcardle_quantum_2020}. The Jordan-Wigner transformation maps each spin orbital to a qubit, whose state $\ket{1}$ or $\ket{0}$ indicates whether the orbital is occupied, and expresses $H$ as a weighted sum of Pauli strings. This allows the energy $E(\boldsymbol{\theta}) = \bra{\psi(\boldsymbol{\theta})} H \ket{\psi(\boldsymbol{\theta})}$ to be estimated and used as $\mathcal{C}$. The circuits are initialised in the Hartree-Fock (HF) state, a single Slater determinant in which the electrons occupy the lowest-energy spin orbitals. Under the Jordan-Wigner mapping, this corresponds to a computational basis state, such as $\ket{1100}$ for two electrons in four spin orbitals.

We use 17 cases of seven molecules in the minimal STO-3G basis~\cite{mcardle_quantum_2020}, summarised in Table~\ref{tab:molecules}. The geometries, Jordan-Wigner Hamiltonians, HF states and exact ground-state energies from full configuration interaction (FCI) are taken from the PennyLane quantum chemistry datasets~\cite{xanadu_quantum_2026}. Each run is judged by its absolute error $|E - E_{\mathrm{FCI}}|$, and an error below $1.6$~mHa, called chemical accuracy~\cite{mcardle_quantum_2020}, is the threshold for success. The HF state treats each electron as moving in the average field of the others, and the part of the energy that it misses, which the circuit has to recover, is the correlation energy $E_{\mathrm{corr}} = E_{\mathrm{HF}} - E_{\mathrm{FCI}}$. For H$_2$, H$_3^+$, H$_4$ and H$_5$, $E_{\mathrm{corr}}$ increases as the bond lengths are stretched, making the stretched hydrogen systems the most strongly correlated and therefore the most challenging cases to simulate accurately~\cite{grimsley_adaptive_2019, mcardle_quantum_2020}.

The operator pool consists of the excitation operators of the unitary coupled-cluster singles and doubles (UCCSD) ansatz~\cite{romero_strategies_2019},
\begin{equation} \label{eq:excitations}
    \tau_{i}^{a} = a_a^\dagger a_i - a_i^\dagger a_a, \qquad
    \tau_{ij}^{ab} = a_a^\dagger a_b^\dagger a_j a_i - a_i^\dagger a_j^\dagger a_b a_a,
\end{equation}
each of which moves one or two electrons from occupied spin orbitals $i, j$ of the HF state to unoccupied spin orbitals $a, b$. We use all single and double excitations that conserve the spin projection, implemented as qubit excitations~\cite{yordanov_qubit-excitation-based_2021}, which omit the sign structure of the Jordan-Wigner mapping,
\begin{equation} \label{eq:qubit_excitations}
    U_{i}^{a}(\theta) = \exp\!\left[\theta \left(Q_a^\dagger Q_i - Q_i^\dagger Q_a\right)\right], \qquad
    U_{ij}^{ab}(\theta) = \exp\!\left[\theta \left(Q_a^\dagger Q_b^\dagger Q_j Q_i - Q_i^\dagger Q_j^\dagger Q_b Q_a\right)\right],
\end{equation}
where $Q_p^\dagger = (X_p - iY_p)/2$ and $Q_p = (X_p + iY_p)/2$ raise and lower qubit $p$ alone. Each excitation carries one parameter, which is initialised from a normal distribution with mean $0$ and standard deviation $0.05$.

In AC-DQAS and C-DQAS, single excitations take the place of the single-qubit gates and double excitations that of the entangling gates. As every single excitation acts on two qubits and every double excitation on four, M-DQAS lists them in two or four transition weight matrices. The repetition penalty is $10^{-4}$~Ha, the SNN receives the operator content (Section~\ref{sec:SNN}), as the energy depends strongly on the angles of the excitations, and the search ends with the BFGS form of the final stage (Section~\ref{sec:DQAS}). Other hyperparameters can be found in Table~\ref{tab:hyperparameters}. As the greedy baseline, ADAPT-VQE (Section~\ref{sec:ADAPT}) runs on the same operator pool, starting from the HF state.

\begin{table}[t]
    \centering
    \caption{Molecular benchmark cases. $n$ is the number of qubits needed to simulate each molecule and pool is the number of single and double excitations of the HF state.}
    \label{tab:molecules}
    \renewcommand{\arraystretch}{1.1}
    \begin{tabular}{lccc}
        \toprule
        \textbf{Molecule} & \textbf{Bond lengths (\AA)} & $n$ & \textbf{Pool (singles + doubles)} \\
        \midrule
        H$_2$ & 0.86, 1.3, 1.7 & 4 & 2 + 1 \\
        HeH$^+$ & 0.86, 1.3, 1.7 & 4 & 2 + 1 \\
        H$_3^+$ & 0.874, 1.3, 1.7 & 6 & 4 + 4 \\
        H$_4$ & 0.5, 0.86, 1.3 & 8 & 8 + 18 \\
        H$_5$ & 0.5, 0.9, 1.5 & 10 & 12 + 42 \\
        LiH & 1.53 & 12 & 16 + 76 \\
        BeH$_2$ & 1.33 & 14 & 24 + 180 \\
        \bottomrule
    \end{tabular}
\end{table}

\section{Results} \label{sec:Results}
We now turn to the results across the three benchmarks, examining how the proposed methods perform in each setting. All simulations were performed on classical hardware using the IRIDIS High Performance Computing Facility at the University of Southampton and PennyLane software~\cite{bergholm_pennylane_2022}. The hyperparameters for all runs are shown in Appendix~\ref{app:Params}. For each method, the main text reports only the exponent $k$ of a power law $c\,n^k$ fitted to the mean circuit evaluations per size $n$. The full fits, with their standard errors, are given in Appendix~\ref{app:Evaluations} and the run times in Appendix~\ref{app:Time}.

\subsection{Unitary Approximation Results}

\begin{figure}[t]
    \centering
    \includegraphics[width=\linewidth]{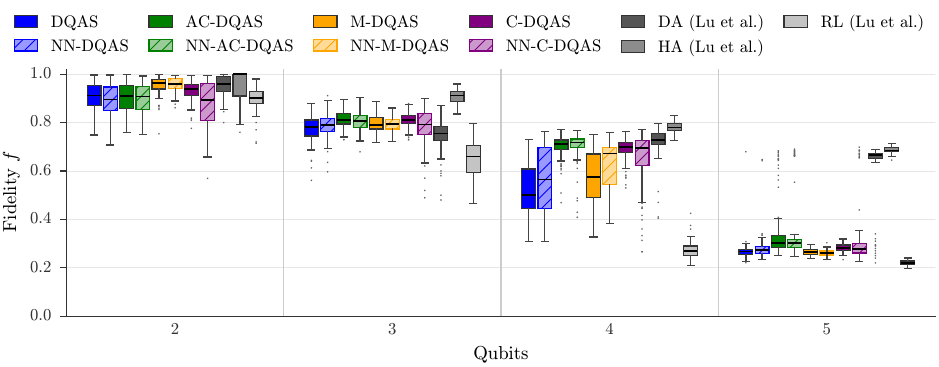}
    \caption{Fidelity $f$ of the final circuits on the 100 QAS-Bench targets of each size. SNN variants are hatched in the colour of their method. Grey are the differentiable (DA), hybrid (HA) and reinforcement-learning (RL) methods evaluated by Lu et al.~\cite{lu_qas-bench_2023}.}
    \label{fig:unitary_f}
\end{figure}

\begin{table}[!t]
    \centering
    \caption{Unitary approximation: median circuit evaluations of the search and gates in the final circuit on the 100 targets of each size $n$. $k$ shows the exponent of a power law $c\,n^k$ fitted to the mean evaluations per size; Table~\ref{tab:fits} gives the standard error of each exponent.}
    \label{tab:unitary}
    \small
    \setlength{\tabcolsep}{8pt}
    \begin{tabular}{lccccccccc}
        \toprule
        & \multicolumn{4}{c}{Evaluations} & & \multicolumn{4}{c}{Gates} \\
        \cmidrule(lr){2-5} \cmidrule(lr){7-10}
        Method & $n = 2$ & 3 & 4 & 5 & $k$ & 2 & 3 & 4 & 5 \\
        \midrule
        DQAS & 37.8k & 53.9k & 57.5k & 60.4k & $0.48$ & 20 & 45 & 80 & 125 \\
        NN-DQAS & 6.7k & 8.1k & 8.7k & 9.5k & $0.31$ & 20 & 45 & 80 & 125 \\
        \addlinespace
        AC-DQAS & 37.0k & 50.9k & 51.2k & 52.4k & $0.40$ & 21 & 50 & 76 & 111 \\
        NN-AC-DQAS & 3.0k & 4.0k & 2.2k & 2.2k & $-0.31$ & 18 & 40 & 71 & 109 \\
        \addlinespace
        M-DQAS & 39.2k & 40.4k & 39.4k & 37.0k & $0.00$ & 20 & 45 & 80 & 125 \\
        NN-M-DQAS & 3.1k & 3.2k & 3.4k & 6.7k & $0.64$ & 20 & 45 & 80 & 125 \\
        \addlinespace
        C-DQAS & 36.7k & 47.9k & 31.1k & 28.5k & $-0.07$ & 21 & 49 & 78 & 112 \\
        NN-C-DQAS & 2.7k & 2.9k & 2.2k & 2.2k & $-0.21$ & 18 & 40 & 71 & 110 \\
        \bottomrule
    \end{tabular}
\end{table}

Figure~\ref{fig:unitary_f} shows the fidelity $f$ achieved on the unitary benchmark, as defined in Eq.~\ref{eq:f}, while Table~\ref{tab:unitary} summarises the search cost and the size of the resulting circuits. From 4 qubits on, AC-DQAS, NN-AC-DQAS, C-DQAS and NN-C-DQAS are the most accurate methods, with AC-DQAS and its variant ahead at 5 qubits. The accuracy of every method falls quickly with the number of qubits, and at 5 qubits all mean fidelities lie between 0.26 and 0.35. 

Compared with the median results reported by Lu et al.~\cite{lu_qas-bench_2023}, AC-DQAS and C-DQAS achieve higher fidelity than RL at every system size, and match DA and HA on smaller systems. At 5 qubits, however, DA and HA achieve median fidelities of 0.67 and 0.69, respectively, roughly twice that of AC-DQAS. Since Lu et al. do not report the circuit sizes produced by their methods, a direct comparison is difficult. To investigate the effect of circuit size, we increased the number of blocks in AC-DQAS from $3n$ to $4n$. This raised its median fidelity at 5 qubits to 0.67, matching DA, while increasing the median circuit size from 111 to 141 gates. This suggests that much of the performance gap at 5 qubits can be attributed to the circuit size. Despite this, we focused on smaller circuits, as our main objective is to compare the performance of our novel methods against each other.

The evaluations of all methods hardly grow with the number of qubits; the fitted exponents lie between $-0.3$ and $0.6$ (Appendix~\ref{app:Evaluations}). From 2 to 5 qubits, the mean evaluations of every method change by at most a factor of 1.9. At 4 and 5 qubits, the median evaluations of C-DQAS are 39 to 46\% lower than those of AC-DQAS, at a median fidelity of 0.70 against 0.71 at 4 qubits and 0.28 against 0.30 at 5 qubits. The surrogate reduces the evaluations by a factor of 5.5 to 24, depending on the method and size, while the mean fidelity of each variant stays within 0.05 of that of its method. These results support the use of NN-C-DQAS, which achieves strong performance while substantially reducing the number of circuit evaluations, with the evaluation cost remaining roughly constant as system size increases. Among the SNN variants, the only exception is NN-M-DQAS, whose evaluations roughly double from 4 to 5 qubits (a median of 3.4k against 6.7k), while those of M-DQAS stay nearly constant. The same saving was found with the cost evaluated on the training state vectors instead of the matrix difference (Appendix~\ref{app:StateVector}). 

At 4 and 5 qubits, AC-DQAS and C-DQAS return fewer gates than DQAS and M-DQAS, which fill all $5n^2$ layers. This is a further advantage of including the architecture constraint, which allows the circuit size to better adapt to the problem.

\subsection{Max-Cut Results}
\begin{figure}[!t]
    \centering
    \includegraphics[width=\linewidth]{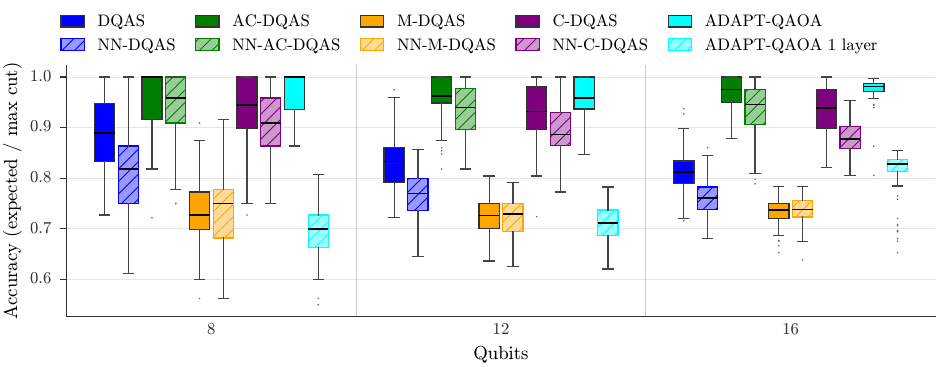}
    \caption{Accuracy (expected cut of the final state divided by the maximum cut) on the 100 Max-Cut graphs of each size. ADAPT-QAOA is shown as published (solid) and stopped after its first layer (hatched).}
    \label{fig:maxcut_accuracy}
\end{figure}

\begin{table}[!t]
    \centering
    \caption{Max-Cut: median circuit evaluations of the search and gates in the final circuit on the 100 graphs with $n$ nodes, and the exponent $k$ of a power law $c\,n^k$ fitted to the mean evaluations. The gates include the $n$ Hadamard gates of the initial state $\ket{+}^{\otimes n}$. ADAPT-QAOA is given as published and, in the last row, stopped after its first layer (Section~\ref{sec:ADAPT}).}
    \label{tab:maxcut}
    \small
    \setlength{\tabcolsep}{8pt}
    \begin{tabular}{lcccccccc}
        \toprule
        & \multicolumn{3}{c}{Evaluations} & & \multicolumn{3}{c}{Gates} \\
        \cmidrule(lr){2-4} \cmidrule(lr){6-8}
        Method & $n = 8$ & 12 & 16 & $k$ & 8 & 12 & 16 \\
        \midrule
        DQAS & 25.0k & 27.7k & 28.0k & $0.19$ & 48 & 72 & 96 \\
        NN-DQAS & 2.6k & 2.6k & 2.5k & $-0.01$ & 48 & 72 & 96 \\
        \addlinespace
        AC-DQAS & 23.0k & 24.9k & 27.3k & $0.29$ & 40 & 58 & 75 \\
        NN-AC-DQAS & 2.5k & 2.8k & 2.9k & $0.20$ & 41 & 60 & 77 \\
        \addlinespace
        M-DQAS & 17.9k & 17.7k & 17.3k & $-0.09$ & 48 & 72 & 96 \\
        NN-M-DQAS & 3.1k & 3.2k & 3.1k & $-0.02$ & 48 & 72 & 96 \\
        \addlinespace
        C-DQAS & 22.5k & 24.8k & 27.6k & $0.32$ & 40 & 59 & 76 \\
        NN-C-DQAS & 2.1k & 2.1k & 2.1k & $-0.01$ & 43 & 63 & 81 \\
        \midrule
        ADAPT-QAOA & 1.1k & 4.3k & 36.8k & $4.76$ & 310 & 1227 & 9445 \\
        ADAPT-QAOA, 1 layer & 141 & 297 & 531 & $1.94$ & 55 & 116 & 212 \\
        \bottomrule
    \end{tabular}
\end{table}

Figure~\ref{fig:maxcut_accuracy} shows the accuracy on the 100 graphs of each size, given by the expected cut of the final state divided by the maximum cut, and Table~\ref{tab:maxcut} the cost of the search and the size of the final circuits. As in the first benchmark, AC-DQAS is the most accurate differentiable method at every size, with at least half of its runs finding the maximum cut exactly at 8 nodes. C-DQAS follows, achieving higher accuracy than DQAS and M-DQAS while requiring a similar number of circuit evaluations to AC-DQAS and producing smaller circuits than the other two methods.

ADAPT-QAOA as published reaches a similar mean accuracy, but with far larger circuits, as every layer contains the full cost Hamiltonian, whereas the standard differentiable methods use at most $6n$ gates. To match the mean accuracy of AC-DQAS, it needs 8 times as many gates at 8 nodes and 94 times as many at 16 nodes (Appendix~\ref{app:AdaptGates}). ADAPT-QAOA evaluates the fewest circuits at 8 nodes and fewer than the standard differentiable methods at 12 nodes, but its count grows as $n^{4.8}$, and at 16 nodes exceeds that of every differentiable method, whose counts are nearly independent of the graph size (fitted exponents $-0.09$ to $0.32$).

The surrogate reduces the circuit evaluations by a factor of 5.5 to 13.1 at every size, while the mean accuracy falls by up to 0.07. Nevertheless, NN-AC-DQAS and NN-C-DQAS remain more accurate than DQAS at every size. Simulation times increase rapidly with graph size. ADAPT-QAOA is faster on 8 and 12 nodes, but takes a similar time to DQAS at 16 nodes. Its run time scales as $n^{10.7}$, compared with $n^{2.0}$ to $n^{4.0}$ for the differentiable methods (Appendix~\ref{app:Time}).

\subsection{Molecular Ground State Results}
\begin{figure}[!t]
    \centering
    \includegraphics[width=\linewidth]{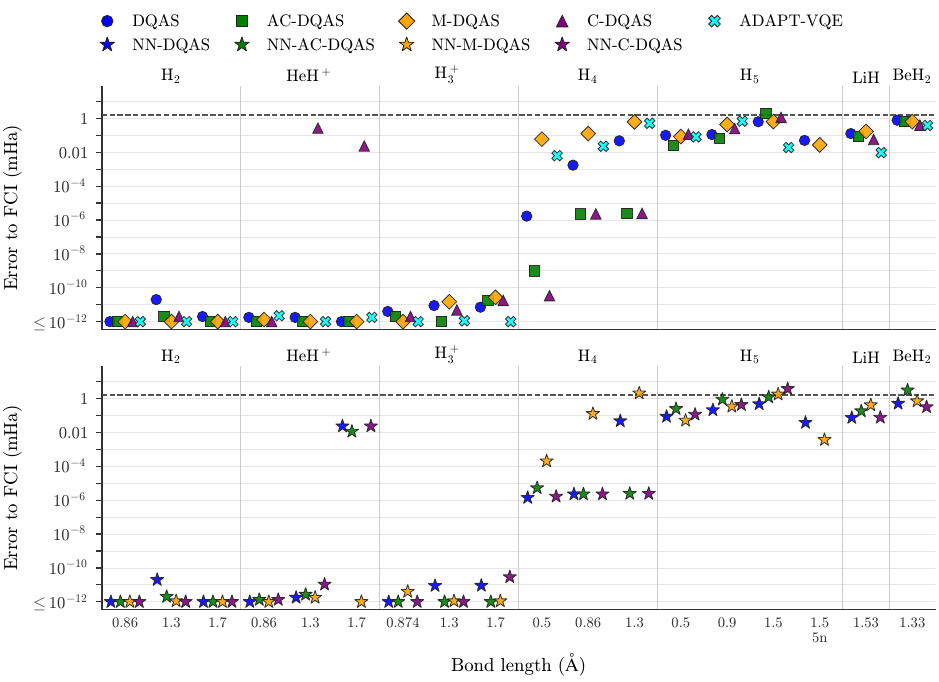}
    \caption{Error of the final energy relative to FCI on each molecular case: the four differentiable methods and ADAPT-VQE (top) and their SNN variants (bottom). Each differentiable method shows the best of its three runs. The slot marked 5n is H$_5$ at 1.5~\AA{} with $L = 5n$. Errors below $10^{-12}$~mHa are capped at the bottom of the axis, and the dashed line marks chemical accuracy (1.6~mHa).}
    \label{fig:chem_error}
\end{figure}

\begin{table}[t]
    \centering
    \caption{Molecular ground states: median circuit evaluations of the search and operators in the final circuit over all runs on the molecules with $n$ qubits (Table~\ref{tab:molecules}), and the exponent $k$ of a power law $c\,n^k$ fitted to the mean evaluations. The H$_5$ runs with $L = 5n$ are excluded.}
    \label{tab:chem}
    \small
    \setlength{\tabcolsep}{6pt}
    \resizebox{\ifdim\width>\linewidth\linewidth\else\width\fi}{!}{%
    \begin{tabular}{lccccccccccccc}
        \toprule
        & \multicolumn{6}{c}{Evaluations} & & \multicolumn{6}{c}{Operators} \\
        \cmidrule(lr){2-7} \cmidrule(lr){9-14}
        Method & $n = 4$ & 6 & 8 & 10 & 12 & 14 & $k$ & 4 & 6 & 8 & 10 & 12 & 14 \\
        \midrule
        DQAS & 22.8k & 25.6k & 42.2k & 55.9k & 33.8k & 31.9k & $0.36$ & 11 & 23 & 32 & 40 & 47 & 56 \\
        NN-DQAS & 2.9k & 4.2k & 5.1k & 7.0k & 7.4k & 8.0k & $0.80$ & 5 & 18 & 32 & 40 & 47 & 56 \\
        \addlinespace
        AC-DQAS & 15.9k & 15.6k & 30.2k & 59.4k & 36.1k & 31.4k & $0.72$ & 4 & 15 & 38 & 49 & 48 & 64 \\
        NN-AC-DQAS & 2.9k & 3.4k & 3.8k & 4.9k & 4.1k & 4.2k & $0.27$ & 3 & 5 & 35 & 47 & 45 & 55 \\
        \addlinespace
        M-DQAS & 15.6k & 18.4k & 31.1k & 48.6k & 33.7k & 35.2k & $0.76$ & 16 & 24 & 32 & 40 & 48 & 56 \\
        NN-M-DQAS & 3.8k & 4.4k & 6.5k & 9.8k & 6.1k & 8.4k & $0.60$ & 16 & 24 & 32 & 40 & 48 & 56 \\
        \addlinespace
        C-DQAS & 15.6k & 26.6k & 53.1k & 80.8k & 39.9k & 44.3k & $0.83$ & 2 & 17 & 38 & 49 & 50 & 67 \\
        NN-C-DQAS & 2.7k & 3.4k & 3.9k & 6.4k & 4.1k & 4.9k & $0.47$ & 2 & 9 & 32 & 50 & 51 & 65 \\
        \midrule
        ADAPT-VQE & 33 & 52 & 736 & 2.7k & 3.6k & 9.3k & $4.96$ & 2 & 2 & 14 & 28 & 28 & 36 \\
        \bottomrule
    \end{tabular}%
    }
\end{table}

Figure~\ref{fig:chem_error} shows the best of the three runs of each method on every molecule and Table~\ref{tab:chem} the cost of all runs and the size of their circuits. Up to 8 qubits, DQAS, AC-DQAS, C-DQAS and ADAPT-VQE reach chemical accuracy in every run and M-DQAS in all but one. On the molecules of 10 to 14 qubits, DQAS reaches chemical accuracy in every run, while AC-DQAS has the lowest median error of the differentiable methods. Every miss on these molecules is on H$_5$ at 1.5~\AA{}, the case with the largest correlation energy, which needs large circuits. ADAPT-VQE uses 49 operators for it, and with $L = 5n = 50$, DQAS and M-DQAS reach chemical accuracy as well. AC-DQAS and C-DQAS learn the size of each block. This allows them to use shallower circuits for H$_2$ and HeH$^+$, while filling most of their $5n$ positions only for the larger molecules. However, for H$_5$ at 1.5~\AA{}, they still narrowly miss chemical accuracy in five of the six runs.

ADAPT-VQE reaches chemical accuracy on every molecule with fewer evaluations than the standard differentiable methods. Its evaluations, however, grow steeply with the size of the molecule ($n^{5.0}$), as every iteration computes the gradient of every pool operator, whereas those of the standard differentiable methods grow as $n^{0.4}$ to $n^{0.8}$. The fitted power laws cross at about 19 to 21 qubits; however, this has not been experimentally verified. The mean evaluations grow by a factor of 1.3 to 2.5 from 4 to 14 qubits for the differentiable methods, against 275 for ADAPT-VQE. Overall, this motivates testing differentiable methods on larger systems where the superior scaling could be realised.

The surrogate reduces the number of circuit evaluations required by the search by a factor of 4.7 to 9.1 on this benchmark, and all four variants evaluate fewer circuits than ADAPT-VQE for BeH$_2$. However, the SNN variants also suffer a loss in accuracy, particularly for more correlated molecules, reaching chemical accuracy in 181 of 210 runs against 202 for the standard methods (Figure~\ref{fig:chem_error}). Nevertheless, NN-DQAS nearly matches DQAS, reaching chemical accuracy in 53 of its 54 runs, including all three on BeH$_2$, where it needs fewer evaluations than ADAPT-VQE (a median of 8.0k against 9.3k).

\section{Discussion}    \label{sec:Discussion}
Across the three benchmarks, AC-DQAS generally provided the strongest performance among the differentiable methods, followed by C-DQAS, DQAS and M-DQAS. This highlights the importance of the alternating-layer constraint, which was particularly effective for unitary approximation and Max-Cut, where the operator pools consist of generic hardware gates. By enforcing single-qubit rotations between entangling gates, the constraint provides a useful structure that an unconstrained search would otherwise have to learn from its samples. The additional gate memory in C-DQAS mainly lowers the cost on unitary approximation and Max-Cut, at some loss in accuracy: on the larger unitaries it needs 39 to 46\% fewer circuit evaluations than AC-DQAS and less time (a median of 26 instead of 42 minutes at 5 qubits), and on the 16-node graphs a similar number of evaluations in less time (91 instead of 103 minutes). For molecular problems, where the operators already conserve the number of electrons, the choice of structure had a smaller impact, and all four methods reached chemical accuracy in most runs (202 of 210). Overall, we recommend AC-DQAS as the default, with C-DQAS as a cheaper alternative when the operator pool consists of generic gates, and DQAS as a more reliable alternative when the pool is tailored to the problem, as for molecules.

An important distinction between the methods is how they handle circuit size. AC-DQAS and C-DQAS learn the size of each block during the search, allowing them to use slightly shallower circuits when sufficient and allocate more gates to harder problems (on the 16-node graphs 75 and 76 gates against 96 for DQAS). In contrast, DQAS and M-DQAS use the full predefined circuit structure, regardless of whether all its positions are needed, which can result in unnecessarily large circuits for simpler problems. This is problematic for NISQ devices, where circuit depth is limited by hardware noise. The ability to adapt the circuit size is therefore an important advantage of the architecture-constrained approaches. For all four methods, however, the maximum circuit size must be chosen in advance, and when it is too restrictive, as for the 5-qubit unitaries and H$_5$ at 1.5~\AA{}, it limits the achievable accuracy.

Comparisons with established methods show that the relative performance depends on the benchmark. On Max-Cut, AC-DQAS matches the accuracy of ADAPT-QAOA (mean accuracies of 0.96 to 0.97 against 0.96 to 0.98) while requiring 8 to 94 times fewer gates, and C-DQAS comes close (0.93), making both attractive for hardware with limited circuit depth. On unitary approximation, both methods are competitive with those reported by Lu et al.~\cite{lu_qas-bench_2023} up to 4 qubits, but achieve around half the median fidelity at 5 qubits (0.30 and 0.28 compared with 0.67 and 0.69). This comparison is not entirely direct, as their methods use larger circuits. Indeed, increasing the number of blocks in AC-DQAS allows it to match the accuracy of their differentiable method at 5 qubits (a median fidelity of 0.67 with 141 instead of 111 gates). For molecular problems, ADAPT-VQE remains more accurate at the system sizes considered. However, a key advantage of the differentiable methods is their scaling. Unlike the ADAPT methods, where the number of evaluations depends on the operator pool and its growth with system size, the evaluation cost of differentiable search is primarily determined by the batch size and number of optimisation steps. Consequently, all four differentiable methods exhibit substantially more favourable scaling in circuit evaluations than the ADAPT methods, with fitted exponents of at most 0.8 against 4.8 for ADAPT-QAOA and 5.0 for ADAPT-VQE. The SNN variants already evaluate fewer circuits than ADAPT-VQE for BeH$_2$ (a mean of 6.2k against 9.3k), although only 6 of their 12 runs there reach chemical accuracy.

The surrogate neural networks further reduce the number of circuit evaluations required by the search (by factors of 12.3, 9.3 and 6.7 on unitary approximation, Max-Cut and molecules, respectively). This comes at the cost of some accuracy, although the reduction is generally small. The mean unitary fidelity decreases by at most 0.05, the mean Max-Cut accuracy by at most 0.07, and the number of molecular runs reaching chemical accuracy drops from 202 to 181 out of 210. The effect depends on both the problem and the underlying search method, with the SNN variants of AC-DQAS and C-DQAS generally offering the most favourable balance between accuracy and evaluation cost. The accuracy loss is more pronounced for highly correlated molecules, suggesting that the SNN variants are less suitable for these cases. At 10 to 14 qubits, only 47 of 66 SNN runs reach chemical accuracy, compared with 59 of 66 for the standard methods.

There is no universally best approach, and the choice should depend on the problem and the available resources. For problems with relatively small operator pools, greedy methods such as ADAPT can remain highly efficient. For larger pools, larger systems or when circuit evaluations are the main limitation, differentiable search offers a more favourable scaling, with AC-DQAS and C-DQAS both providing a strong balance between accuracy and circuit size. Adding the surrogate makes these methods considerably more evaluation-efficient, but for particularly challenging problems, such as highly correlated molecular systems, the standard methods are more reliable.

The proposed methods have further limitations. They introduce additional hyperparameters, including $K_{\max}$ and separate learning rates, which need to be selected for each benchmark. Another limitation is that all methods assume all-to-all connectivity, which is difficult to realise on current hardware; as this applies equally to every method compared, including ADAPT, we expect the relative trends to hold. Finally, all results were obtained using noiseless simulations and exact expectation values, meaning that the reported circuit counts do not capture the additional measurement cost that would arise on real quantum hardware.

Future work will explore learning the maximum circuit size during the search, rather than fixing it in advance. We also plan to extend the role of the neural network surrogate beyond ranking candidate circuits. In particular, it could be used to predict promising initial parameters for each candidate, potentially reducing the number of circuit evaluations required during optimisation. Finally, we aim to investigate the robustness of the proposed methods under realistic hardware conditions by introducing noise into the simulations, restricting qubit connectivity, and testing them on real quantum hardware.

In summary, the alternating-layer structure of AC-DQAS and C-DQAS gave the most accurate differentiable methods on most problems. Compared with the established ADAPT methods, the differentiable methods scale much better (circuit evaluations growing as at most $n^{0.8}$ against $n^{4.8}$ to $n^{5.0}$), making them suitable for NISQ computations on larger systems. Furthermore, surrogate neural networks reduce the number of circuit evaluations by a factor of 6.7 to 12.3 while largely maintaining accuracy. Overall, surrogate-assisted differentiable search with extra constraints is a promising route towards QAS within the evaluation budgets of near-term devices.

\begin{acknowledgements}
This work was supported by the Engineering and Physical Sciences Research Council (EPSRC) [grant number EP/Y035267/1] through the Centre for Doctoral Training in Quantum Technology Engineering at the University of Southampton. The authors acknowledge the use of the IRIDIS High Performance Computing Facility, and associated support services at the University of Southampton, in the completion of this work.
\end{acknowledgements}

\section*{Author contributions}
D.K. and T.B. developed the methods. D.K. implemented them, performed the simulations and wrote the manuscript. T.B. supervised the project and revised the manuscript.

\section*{Code and data availability}
The code used in this study, the Max-Cut graphs and the results of every run are available on request from \href{mailto:dk2u25@soton.ac.uk}{dk2u25@soton.ac.uk}. The target unitaries are from QAS-Bench~\cite{lu_qas-bench_2023} and the molecular Hamiltonians are from the PennyLane quantum chemistry datasets~\cite{xanadu_quantum_2026}, both publicly available.

\bibliographystyle{quantum}
\bibliography{references}

\clearpage
\appendix
\section{Experimental Hyperparameters} \label{app:Params}
Table~\ref{tab:hyperparameters} lists the settings of the differentiable methods and the learning rates of their SNN variants. For AC-DQAS and C-DQAS, $L$ counts composite blocks, each consisting of a layer of single-qubit operations followed by a layer of two-qubit operations. For H$_5$ at 1.5~\AA{}, DQAS, M-DQAS and their SNN variants were also run with $L = 5 \times n_{\text{qubits}}$.

\begin{table}[H]
    \centering
    \caption{Hyperparameter configurations across the three benchmarks. $n_{\text{qubits}}$ denotes the number of qubits, and $n_{\text{ops}}^{(2)}$ denotes the number of available two-qubit operations. NN rows show the learning rates of the SNN variant of the method.}
    \label{tab:hyperparameters}
    \renewcommand{\arraystretch}{1.1}
    \resizebox{0.9\textwidth}{!}{
    \begin{tabular}{llccc}
        \toprule
        \textbf{Algorithm} & \textbf{Parameter} & \textbf{Unitary approximation} & \textbf{Max-Cut} & \textbf{Molecules} \\
        \midrule
        \multirow{9}{*}{\textbf{DQAS}}
        & Operator pool $\mathcal{G}$ & $R_X, R_Y, R_Z, \text{CNOT}$ & $R_X, R_Y, R_Z, \text{CNOT}$ & Excitations \\
        & Layers $L$ & $5 \times n_{\text{qubits}}^2$ & $5 \times n_{\text{qubits}}$ & $4 \times n_{\text{qubits}}$ \\
        & Batch size $B$ & $64$ & $64$ & $64$ \\
        & Max steps $S$ & $1000$ & $1000$ & $1000$ \\
        & Final samples $F$ & $16$ & $16$ & Most likely \\
        & Refinement steps $R$ & $300$ & $300$ & BFGS \\
        & Learning rates $\eta_\alpha, \eta_\theta$ & $0.5, 0.05$ & $0.5, 0.05$ & $0.1, 0.01$ \\
        & NN: $\eta_\alpha, \eta_\theta$ & $0.5, 0.05$ & $0.5, 0.05$ & $0.1, 0.01$ \\
        & Min improvement $\varepsilon$ & $10^{-4}$ & $10^{-4}$ & $10^{-4}$ \\
        \midrule
        \multirow{11}{*}{\textbf{AC-DQAS}}
        & 1Q ops $\mathcal{G}_1$ & $R_X, R_Y, R_Z$ & $R_X, R_Y, R_Z$ & Singles \\
        & 2Q ops $\mathcal{G}_2$ & $\text{CNOT}$ & $\text{CNOT}$ & Doubles \\
        & Blocks $L$ & $3 \times n_{\text{qubits}}$ & $3$ & $2$ \\
        & Max 2Q ops per block $K_{\max}$ & $\min(4, n_{\text{ops}}^{(2)})$ & $\min(n_{\text{qubits}}-1, n_{\text{ops}}^{(2)})$ & $\lceil 1.5 \times n_{\text{qubits}} \rceil$ \\
        & Batch size $B$ & $64$ & $64$ & $64$ \\
        & Max steps $S$ & $1000$ & $1000$ & $1000$ \\
        & Final samples $F$ & $16$ & $16$ & Most likely \\
        & Refinement steps $R$ & $300$ & $300$ & BFGS \\
        & Learning rates $\eta_\alpha, \eta_\theta$ & $0.5, 0.05$ & $0.5, 0.05$ & $0.1, 0.01$ \\
        & NN: $\eta_\alpha, \eta_\theta$ & $0.5, 0.05$ & $0.5, 0.05$ & $0.1, 0.005$ \\
        & Min improvement $\varepsilon$ & $10^{-4}$ & $10^{-4}$ & $10^{-4}$ \\
        \midrule
        \multirow{9}{*}{\textbf{M-DQAS}}
        & Operator pool $\mathcal{G}$ & $R_X, R_Y, R_Z, \text{CNOT}$ & $R_X, R_Y, R_Z, \text{CNOT}$ & Excitations \\
        & Layers $L$ & $5 \times n_{\text{qubits}}^2$ & $5 \times n_{\text{qubits}}$ & $4 \times n_{\text{qubits}}$ \\
        & Batch size $B$ & $64$ & $64$ & $64$ \\
        & Max steps $S$ & $1000$ & $1000$ & $1000$ \\
        & Final samples $F$ & $16$ & $16$ & Most likely \\
        & Refinement steps $R$ & $300$ & $300$ & BFGS \\
        & Learning rates $\eta_\beta, \eta_\theta$ & $10^{-4}, 0.05$ & $10^{-3}, 0.05$ & $0.3, 0.01$ \\
        & NN: $\eta_\beta, \eta_\theta$ & $10^{-3}, 0.05$ & $10^{-3}, 0.05$ & $0.1, 0.005$ \\
        & Min improvement $\varepsilon$ & $10^{-4}$ & $10^{-4}$ & $10^{-4}$ \\
        \midrule
        \multirow{11}{*}{\textbf{C-DQAS}}
        & 1Q ops $\mathcal{G}_1$ & $R_X, R_Y, R_Z$ & $R_X, R_Y, R_Z$ & Singles \\
        & 2Q ops $\mathcal{G}_2$ & $\text{CNOT}$ & $\text{CNOT}$ & Doubles \\
        & Blocks $L$ & $3 \times n_{\text{qubits}}$ & $3$ & $2$ \\
        & Max 2Q ops per block $K_{\max}$ & $\min(4, n_{\text{ops}}^{(2)})$ & $\min(n_{\text{qubits}}-1, n_{\text{ops}}^{(2)})$ & $\lceil 1.5 \times n_{\text{qubits}} \rceil$ \\
        & Batch size $B$ & $64$ & $64$ & $64$ \\
        & Max steps $S$ & $1000$ & $1000$ & $1000$ \\
        & Final samples $F$ & $16$ & $16$ & Most likely \\
        & Refinement steps $R$ & $300$ & $300$ & BFGS \\
        & Learning rates $\eta_\alpha, \eta_\theta$ & $10^{-2}, 10^{-2}$ & $0.5, 0.05$ & $0.02, 0.005$ \\
        & NN: $\eta_\alpha, \eta_\theta$ & $0.5, 0.05$ & $0.5, 0.05$ & $0.02, 0.005$ \\
        & Min improvement $\varepsilon$ & $10^{-4}$ & $10^{-4}$ & $10^{-4}$ \\
        \bottomrule
    \end{tabular}
    }
\end{table}

\section{Additional Results} \label{app:Additional}

\subsection{Scaling of the Circuit Evaluations} \label{app:Evaluations}

Table~\ref{tab:fits} reports the median number of circuit evaluations at the largest system size for each benchmark, along with the exponent $k$ and its standard error for each power law fitted to the mean evaluations. On unitary approximation and Max-Cut, the exponents of the differentiable methods are close to zero and their mean evaluations change by at most a factor of 1.9 between any two sizes. On molecules, their evaluations depend more on the molecule than on the number of qubits, with the most at 10 qubits (H$_5$), so their exponents are less certain (standard errors of 0.17 to 0.38). Nevertheless, from the smallest to the largest size, the mean evaluations of ADAPT-QAOA grow by a factor of 29 and those of ADAPT-VQE by 275, against at most 1.3 (Max-Cut) and 2.5 (molecules) for the differentiable methods. This shows the clear benefit of the differentiable methods. Note that the exponents rest on only three sizes for Max-Cut, four for unitary approximation and six for molecules, so they indicate the trend, nearly flat for the differentiable methods and steep for the ADAPT methods, rather than precise scaling laws.

\begin{table}[H]
    \centering
    \caption{Median circuit evaluations of the search at the largest size (unitary approximation 5 qubits, Max-Cut 16 nodes, molecules 14 qubits) and exponent $k$ of a power law $c\,n^k$ fitted to the mean evaluations per size, with its standard error.}
    \label{tab:fits}
    \small
    \setlength{\tabcolsep}{4pt}
    \resizebox{\ifdim\width>\linewidth\linewidth\else\width\fi}{!}{%
    \begin{tabular}{lcccccc}
        \toprule
        & \multicolumn{2}{c}{Unitary approximation} & \multicolumn{2}{c}{Max-Cut} & \multicolumn{2}{c}{Molecules} \\
        \cmidrule(lr){2-3} \cmidrule(lr){4-5} \cmidrule(lr){6-7}
        Method & Evaluations & $k$ & Evaluations & $k$ & Evaluations & $k$ \\
        \midrule
        DQAS & 60.4k & $0.48 \pm 0.08$ & 28.0k & $0.19 \pm 0.03$ & 31.9k & $0.36 \pm 0.28$ \\
        NN-DQAS & 9.5k & $0.31 \pm 0.03$ & 2.5k & $-0.01 \pm 0.04$ & 8.0k & $0.80 \pm 0.17$ \\
        \addlinespace
        AC-DQAS & 52.4k & $0.40 \pm 0.12$ & 27.3k & $0.29 \pm 0.03$ & 31.4k & $0.72 \pm 0.35$ \\
        NN-AC-DQAS & 2.2k & $-0.31 \pm 0.32$ & 2.9k & $0.20 \pm 0.04$ & 4.2k & $0.27 \pm 0.23$ \\
        \addlinespace
        M-DQAS & 37.0k & $0.00 \pm 0.05$ & 17.3k & $-0.09 \pm 0.02$ & 35.2k & $0.76 \pm 0.30$ \\
        NN-M-DQAS & 6.7k & $0.64 \pm 0.26$ & 3.1k & $-0.02 \pm 0.05$ & 8.4k & $0.60 \pm 0.27$ \\
        \addlinespace
        C-DQAS & 28.5k & $-0.07 \pm 0.22$ & 27.6k & $0.32 \pm 0.11$ & 44.3k & $0.83 \pm 0.38$ \\
        NN-C-DQAS & 2.2k & $-0.21 \pm 0.17$ & 2.1k & $-0.01 \pm 0.01$ & 4.9k & $0.47 \pm 0.20$ \\
        \midrule
        ADAPT-QAOA &  &  & 36.8k & $4.76 \pm 0.90$ &  &  \\
        \quad 1 layer &  &  & 531 & $1.94 \pm 0.07$ &  &  \\
        ADAPT-VQE &  &  &  &  & 9.3k & $4.96 \pm 0.75$ \\
        \bottomrule
    \end{tabular}%
    }
\end{table}

\subsection{Run Time} \label{app:Time}

Table~\ref{tab:time} gives the run times of the search, which grow steeply with the number of qubits. The run times of the SNN variants grow more slowly than those of their methods, except for NN-M-DQAS on the unitaries. Once again, the scaling of the differentiable methods is far better than that of the ADAPT methods.

\begin{table}[H]
    \centering
    \caption{Median run time of the search at the largest size (unitary approximation 5 qubits, Max-Cut 16 nodes,
    molecules 14 qubits) and exponent $k$ of a power law $c\,n^k$ fitted to the mean run time per size, with its standard error.}
    \label{tab:time}
    \small
    \resizebox{\ifdim\width>\linewidth\linewidth\else\width\fi}{!}{%
    \begin{tabular}{lcccccc}
        \toprule
        & \multicolumn{2}{c}{Unitary approximation} & \multicolumn{2}{c}{Max-Cut} & \multicolumn{2}{c}{Molecules} \\
        \cmidrule(lr){2-3} \cmidrule(lr){4-5} \cmidrule(lr){6-7}
        Method & Time (min) & $k$ & Time (min) & $k$ & Time (min) & $k$ \\
        \midrule
        DQAS & 41.9 & $2.30 \pm 0.05$ & 122.6 & $3.99 \pm 1.19$ & 58.0 & $3.41 \pm 0.29$ \\
        NN-DQAS & 21.5 & $1.60 \pm 0.05$ & 39.8 & $3.31 \pm 1.28$ & 29.8 & $3.07 \pm 0.27$ \\
        \addlinespace
        AC-DQAS & 42.1 & $2.44 \pm 0.14$ & 103.0 & $3.56 \pm 1.09$ & 55.9 & $3.90 \pm 0.44$ \\
        NN-AC-DQAS & 13.1 & $1.14 \pm 0.32$ & 35.1 & $2.57 \pm 0.80$ & 14.6 & $1.90 \pm 0.27$ \\
        \addlinespace
        M-DQAS & 35.8 & $2.02 \pm 0.01$ & 83.5 & $3.38 \pm 1.15$ & 95.9 & $3.59 \pm 0.35$ \\
        NN-M-DQAS & 135.5 & $2.91 \pm 0.36$ & 89.0 & $1.97 \pm 0.22$ & 148.5 & $2.37 \pm 0.30$ \\
        \addlinespace
        C-DQAS & 25.7 & $1.90 \pm 0.22$ & 91.2 & $3.33 \pm 1.00$ & 85.2 & $3.77 \pm 0.40$ \\
        NN-C-DQAS & 32.4 & $1.37 \pm 0.21$ & 42.6 & $2.25 \pm 0.73$ & 29.7 & $2.01 \pm 0.27$ \\
        \midrule
        ADAPT-QAOA &  &  & 121.9 & $10.65 \pm 2.84$ &  &  \\
        ADAPT-VQE &  &  &  &  & 11.3 & $7.62 \pm 1.01$ \\
        \bottomrule
    \end{tabular}%
    }
\end{table}

\subsection{Circuit Size of ADAPT-QAOA} \label{app:AdaptGates}

\begin{figure}[H]
    \centering
    \includegraphics[width=0.9\linewidth]{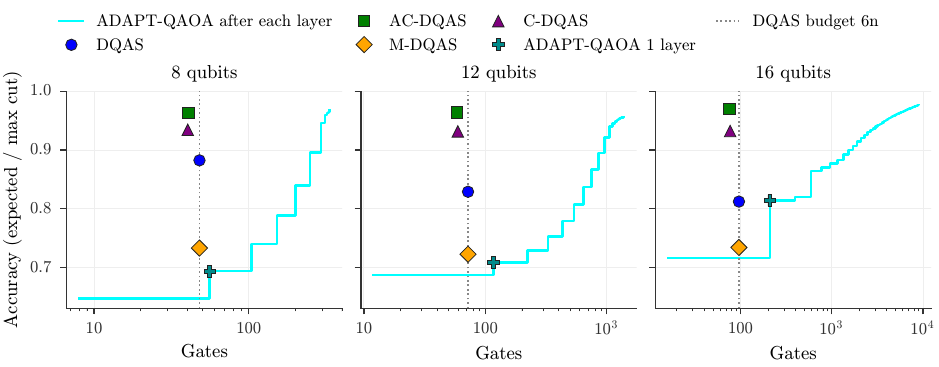}
    \caption{Mean accuracy of ADAPT-QAOA against its mean number of gates after each layer, alongside the mean accuracy and gate count of the differentiable methods; ADAPT-QAOA 1 layer is the first step of the curve. The dotted line indicates the maximum of $6n$ gates allowed for the differentiable methods.}
    \label{fig:maxcut_adapt_gates}
\end{figure}
Figure~\ref{fig:maxcut_adapt_gates} shows the mean accuracy of ADAPT-QAOA as its circuit grows layer by layer. ADAPT-QAOA requires substantially larger circuits than AC-DQAS and C-DQAS to reach comparable accuracy (up to 94 times as many gates). This highlights a key limitation of the method, particularly for near-term hardware, where circuit depth is restricted. Combined with the increasing number of circuit evaluations and optimisation time required as the system size grows, this makes ADAPT-QAOA less suitable for larger problems.

\subsection{Cost Evaluated on Training States} \label{app:StateVector}

To check that the saving of the surrogate does not depend on the matrix cost (Eq.~\ref{eq:matrix_comp}), DQAS and NN-DQAS were also run on the 2-qubit unitaries with the cost evaluated through the fidelity (Eq.~\ref{eq:f}) on the individual training states. NN-DQAS then needed 5.7 times fewer evaluations than DQAS during the search (a median of 0.81 against 4.56 million) at nearly the same mean fidelity (0.756 against 0.761). As every cost evaluation passes 200 training states through the circuit, this experiment was run only at the smallest size, but it shows the same pattern as the matrix cost, where NN-DQAS needs 5.6 times fewer evaluations at 2 qubits (Table~\ref{tab:unitary}).

\end{document}